\documentclass[aps,prc,preprintnumbers,epsf]{revtex4}
\newcommand{\gtap}{\;{\raise.3ex\hbox{$>$\kern-.75em\lower1ex\hbox{$\sim$}}}\;}
\newcommand{\ltap}{\;{\raise.3ex\hbox{$<$\kern-.75em\lower1ex\hbox{$\sim$}}}\;}
\usepackage[dvips]{graphicx}
\newcommand{\bea}{\begin{eqnarray}}
\newcommand{\eea}{\end{eqnarray}}

\begin{document}
\preprint{arXiv: 0808.2850[nucl-th]}
\title{Pad\'e expansion and nucleon-nucleon scattering in coupled
channels}
\author{Dan Liu, Ji-Feng Yang\footnote{corresponding author.}}
\address{Department of Physics, East China Normal University,
Shanghai, 200062, China}
\date{\today}
\begin{abstract}
We extend our Pad\'e-aided analysis of the nonperturbative
renormalization of nucleon-nucleon scattering to the case of coupled
channels.\end{abstract} \maketitle

Since the significant effective field theory (EFT) approach to
nucleon systems\cite{BeK,EpelReview}, there has been creating
controversies about the consistent renormalization and EFT power
counting in nonperturbative regime\cite{NTvK,Epel}. The main
difficulty is due to the nontrivial prescription dependence
developed in nonperturbative regime, where some wisdoms about
renormalization established within perturbative regimes cease to
apply directly\cite{C71}. Therefore, the nonperturbative
prescription dependence must be removed through imposing appropriate
boundary conditions which are usually implemented through various
forms of data fitting either explicitly or
implicitly\cite{vK,EGM,Rho,Fred,Gege,BBSvK,Soto,VA}. In more recent
literature, nonperturbative counter terms, or equivalently,
nonperturbative parametrization of the renormalization prescription
has been the main focus\cite{YPhillips,EPVRAM}. That is, due to the
difficulty in treating the issues in nonperturbative regimes, the
key issue is to find more efficient parametrization of
nonperturbative prescription.

In this regard, we have performed an analysis and treatment of the
nonperturbative prescription dependence basing on Pad\'e approximant
of an important nonperturbative factor of $T$-matrix for
nucleon-nucleon ($NN$) scattering in uncoupled channles\cite{
g_npt,CTP47,CPL23}. In this report, we will extend our analysis to
coupled channels. Some general theoretical and technical issues
associated with coupled channels will be addressed first, then we
will illustrate our method in $^3D_3$$-$$^3G_3$.

The object under consideration is the $T$-matrix for
nucleon-nucleon scattering processes at low energies. According to
Weinberg's proposal\cite{WeinEFT}, this $T$-matrix should be solved
from Lippmann-Schwinger (LS) equation with the potential to be
systematically constructed using $\chi$PT as the low energy
effective theory of QCD. For coupled channels, such LSE's read,
\begin{eqnarray}
\label{LSE} &&{\bf T}(p^{\prime},p; E)={\bf
V}(p^{\prime},p;E)+\int_k G_0(k;E^+){\bf V}(p^{\prime},k
;E) \times {\bf T}(k,p; E), \nonumber \\
&&G_0(k;E^+)\equiv \frac{1}{E^{+}-k^2/M},\ \ E^{+}\equiv E + i
\epsilon,
\end{eqnarray}
with $E$ being the nucleon energy in the center mass frame, $M$ the
nucleon mass, $p^{\prime}=|{\bf p}^\prime|$, $p=|{\bf p}|$. The
bold-faced capital letters represent the $2\times2$ matrix-valued
objects in the angular quantum number space. The convolution is
understood as already regularized and/or renormalized in an
unspecified prescription in order to make our discussions generally
valid. Following Refs.\cite{g_npt,CTP47,CPL23}, the above LSE for
coupled channels could be transformed into a compact and hence
nonperturbative parametrization of $T$-matrix as below,
\begin{eqnarray}
\label{npt} && \ {\bf T}^{-1}(p^{\prime},p; E)={\bf
V}^{-1}(p^{\prime},p;E)-{\mathcal{G}}(p^{\prime},p;
E),\\
&&\label{g-factor}{\mathcal{G}}(p^{\prime},p; E)\equiv {\bf
V}^{-1}(p^{\prime},p;E)\times \left[\int_k G_0(k;E^+) {\bf
V}(p^{\prime},k;E )\times {\bf T}(k,p; E)\right]\times {\bf
T}^{-1}(p^{\prime},p; E)
\end{eqnarray}where the factor ${\mathcal{G}}$
assumes all the 'loop' processes generated by $\bf V$ in the
field-theoretical terminology. Making use of the $K$-matrix
formalism, the unitarity of such compact $T$-matrices follows
immediately\cite{g_npt,CTP47,CPL23}. It is also easy to verify the
inverse relation in the coupled channels: \bea && {\bf T}\times {\bf
T}^{-1}=\left({\bf V}+\int G_0{\bf V}\times {\bf T}\right)\times{\bf
T}^{-1}={\bf V}\times{\bf T}^{-1}+ {\bf V}\times{\mathcal{G}}= {\bf
V}\times({\bf T}^{-1}+{\mathcal{G}})={\bf V}\times{\bf V}^{-1}={\bf I},\\
&&{\bf T}^{-1}\times{\bf T}=({\bf V}^{-1}-{\mathcal{G}})\times{\bf
T})={\bf V}^{-1}\times{\bf T}-{\mathcal{G}}\times{\bf T}={\bf
V}^{-1}\times \left ({\bf T}-\int G_0 {\bf V}\times {\bf T}\right
)={\bf V}^{-1}\times{\bf V}={\bf I},\eea with ${\bf I}$ denoting the
$2\times2$ unit matrix. Obviously, in terms of $\bf V$ and
${\mathcal{G}}$, $\bf T$ are nonperturbative objects.

It is interesting to note that in terms of the standard
parametrization of $S$-matrix, \bea\label{sm-phase} {\bf S}=
\left(%
\begin{array}{cc}
  \cos2\epsilon_j(p) \exp[2i\delta^{1j}_{j-1}(p)] &
  i\sin2\epsilon_j(p)
  \exp[i(\delta^{1j}_{j-1}(p)+\delta^{1j}_{j+1}(p))] \\
  i\sin2\epsilon_j (p)  \exp[i(\delta^{1j}_{j+1}(p)+
  \delta^{1j}_{j-1}(p))]  &
  \cos2\epsilon_j (p)\exp[2i\delta^{1j}_{j+1}(p)]\\
\end{array}%
\right)={\bf I}-i\frac{Mp}{2\pi}{\bf T},\eea we could find that,
\bea \label{INVT} {\bf T}^{-1}=i\frac{Mp}{4\pi}{\bf
I}+\frac{Mp}{4\pi}
\left(%
\begin{array}{cc}
\frac{\sin(\delta^{1j}_{j-1}+\delta^{1j}_{j+1})-
\sin(\delta^{1j}_{j-1}
-\delta^{1j}_{j+1})\cos(2\epsilon_j)}{\cos(\delta^{1j}_{j-1}+
\delta^{1j}_{j+1}) -\cos(\delta^{1j}_{j-1}
-\delta^{1j}_{j+1})\cos(2\epsilon_j)},&
\frac{-\sin(2\epsilon_j)}{\cos(\delta^{1j}_{j-1}+\delta^{1j}_{j+1})
-\cos(\delta^{1j}_{j-1} -\delta^{1j}_{j+1})\cos(2\epsilon_j)}
   \\
\frac{-\sin(2\epsilon_j)}{\cos(\delta^{1j}_{j-1}+
\delta^{1j}_{j+1}) -\cos(\delta^{1j}_{j-1}
-\delta^{1j}_{j+1})\cos(2\epsilon_j)}, &
\frac{\sin(\delta^{1j}_{j-1}+\delta^{1j}_{j+1})-
\sin(\delta^{1j}_{j+1}
-\delta^{1j}_{j-1})\cos(2\epsilon_j)}{\cos(\delta^{1j}_{j-1}+
\delta^{1j}_{j+1}) -\cos(\delta^{1j}_{j+1}
-\delta^{1j}_{j-1})\cos(2\epsilon_j)}\\
\end{array}%
\right).\eea Thus, the imaginary part of ${\bf T}^{-1}$ is simple
and proportional to a unit matrix, the off-diagonal entries of ${\bf
T}^{-1}$ are real numbers only. The unitarity now reads: $ {\bf
T}^{-1}-({\bf T}^{-1})^*=i\frac{Mp}{2\pi}{\bf I}$. Only the real
part of ${\bf T}^{-1}$ (or ${\mathcal{G}}$) is subject to
nonperturbative renormalization and hence the unitarity is not
affected by renormalization. The prescription dependence is
exclusively contained in the real part of the factor
${\mathcal{G}}$.

The motivations and plausibility of employing Pad\'e approximant to
${\mathcal{G}}$ were already demonstrated in
Refs.\cite{g_npt,CTP47,CPL23}. In this approximation, the unitarity
of $T$-matrices is automatically preserved, a virtue that is welcome
in hadron physics\cite{ZhengHQ}. Another important virtue is its
generality and flexibility in parametrizing the prescription
dependence, avoiding being stuck in or confined to a special
prescription that might not be quite compatible with physical
boundaries.

Now it is clear that ${\mathcal{G}}$ is a $2\times2$ matrix with the
diagonal entries being complex, while the off-diagonal ones real.
The Pad\'e approximant is applied to the real part of any matrix
element, diagonal or off-diagonal ($p=\sqrt{ME}$):
\begin{eqnarray}
\text{Re}\{{\mathcal{G}}(p)_{l+\Delta,l+\Delta^\prime}\}|_{\text{Pad\'e}}
=\frac{\sum_k N_{l+\Delta,l+\Delta^\prime;k}p^{2k}}{\sum_{k^\prime}
D_{l+\Delta,l+\Delta^\prime;k^\prime}p^{2k^\prime}},\ \
\Delta,\Delta^\prime=0,2.
\end{eqnarray}Evidently, in Pad\'e approximant, the prescription
dependence is contained the parameters $[N_{\cdots;k},
D_{\cdots;k^\prime}]$. In other words, $[N_{\cdots;k},
D_{\cdots;k^\prime}]$ serve as approximate parametrization of the
nonperturbative prescription.

Then the renormalized $\bf T$ could be approximately parametrized in
nonperturbative regime as follows,\bea\label{pade_T}&& {\bf
T}_{\text{\tiny
Pad\'e}}^{-1}(p;[g_{\cdots},C_{\cdots}];[N_{\ldots},D_{\ldots}])
={\bf V}^{-1}(p;[g_{\cdots},C_{\cdots}])- \left[\frac{\sum_k
N_{l+\Delta,l+\Delta^\prime;k}p^{2k}}{\sum_{k^\prime}
D_{l+\Delta,l+\Delta^\prime;k^\prime}p^{2k^\prime}}\right]_{2\times2}+
i \frac{Mp}{4\pi}{\bf I}.\eea Simple and coarse as it is, such
nonperturbative parametrization of the $T$-matrix contains all
contributing parameters: EFT couplings $[g_{\cdots},C_{\cdots}]$,
and renormalization prescription parameterized in terms of
$[N_{\cdots;k}, D_{\cdots;k^\prime}]$. As a byproduct, the Pad\'e
parameters allows us in principle to effectively imitate any
renormalization prescription of $\bf T$ through corresponding
definition of $[N_{\cdots;k}, D_{\cdots;k^\prime}]$.

For our approximation to be sensible, the Pad\'e parameters must be
appropriately determined. Then their magnitude orders should be in
accordance with EFT power counting. For a general EFT power
counting, one may expect that: \bea\label{pade-PC}
\left\{\frac{N_{\cdots;i}}{N_{\cdots;0}},
\frac{D_{\cdots;i}}{D_{\cdots;0}}\right\}\sim
\Lambda^{F(i)}\mu^{f(i)},\eea with $\Lambda(\sim 500\text{MeV})$
being the upper EFT scale, $\mu$ the typical EFT scale, here, say,
$\sim(10, 100)\text{MeV}$. $F, f$ are some counting functions. Note
that the reflection of EFT power counting in the factor
${\mathcal{G}}$ is completely nonperturbative, in sheer contrast to
the conventional understandings.

Occasional large deviation from such rules should be due to
unnatural behaviors of the $NN$ scattering. The EFT approach would
be indeed problematic only if no power counting scheme could be
sensibly realized in any renormalization prescription. In other
words, the failure of some power counting schemes does not imply the
very failure of the EFT approach.

In general, the intrinsic scales involved in the Pad\'e parameters
should be $\Lambda$ and $\mu$: \bea\label{Pade-scale}
\left|\frac{N_{\cdots;0}}{N_{\cdots;k}}\right|^{\frac{1}{2k}}
\text{or}\left|\frac{D_{\cdots;0}}{D_{\cdots;k}}
\right|^{\frac{1}{2k}}\sim\mu^\alpha\Lambda^{1-\alpha},\  \
\alpha\in (0,1.0), \ k>0.\eea As $\text{dim}[{\mathcal{G}}]=2$, we
choose $\text{dim}[D_{\cdots;0}]=2$ and hence
$|D_{\cdots;0}|^{\frac{1}{2}}\sim \mu^\alpha\Lambda^{1-\alpha},\
\alpha\in (0,1.0)$. This is because Pad\'e parameters are in fact
functions of both EFT couplings and renormalization scales or
constants.

Now, to explore physics using the parametrization given in
Eq.(\ref{pade_T}), we must fix the prescription or Pad\'e parameters
through imposing appropriate boundary conditions. To this end, as in
most literature, we fit to the PWA\cite{PWA} data for the phase
shifts and mixing angle in the low energy ends, say the kinetic
laboratory energy $T_{\text{Lab}}(=2E)\in (0, 50)\text{MeV}$. In
coupled channels, one must fit three sets of Pad\'e parameters for
the phase shifts and mixing angle at the same time. In order to see
the main points or rationalities of the Pad\'e-aided analysis of
coupled channels, we will work with the simplest and hence coarsest
cases of Pad\'e approximant, i.e., the constant ${\mathcal{G}}$
factor,
\bea\label{constPade}\text{Re}({\mathcal{G}}(p))|_{\text{Pad\'e}}
\approx \left(\begin{array}{cc} g^0_{j-1,j-1} & g^0_{j-1,j+1} \\
                   g^0_{j-1,j+1} & g^0_{j+1,j+1} \\
                   \end{array}
                   \right).
\eea Alternatively, with such choice of pad\'e approximant, we wish
to probe the most important scales in the nonperturbative factors
${\mathcal{G}}$, in order to see if there would be significant
deviation from the EFT power counting, or abnormal numbers.

In this short report, we pick up the $^3D_3$$-$$^3G_3$ channels for
illustration where up to next-to-next-to-leading order the
potentials contain no extra contact terms to be determined first.
This is also the highest coupled channels where at least one
channel, $^3D_3$, is not perturbative, which could be seen below. As
before, we employ the potentials and couplings given by
EGM\cite{EGM}. One could well employ other sets of definitions for
comparison. Such works will be carried out in the future. We also
note that we deliberately work with low precision in order to save
computer workloads: round up to the first two digits.

\begin{figure}[t]
  \begin{center}
    \begin{tabular}{ccc}
  \hspace*{-0.75cm}\resizebox{58mm}{!}{\includegraphics{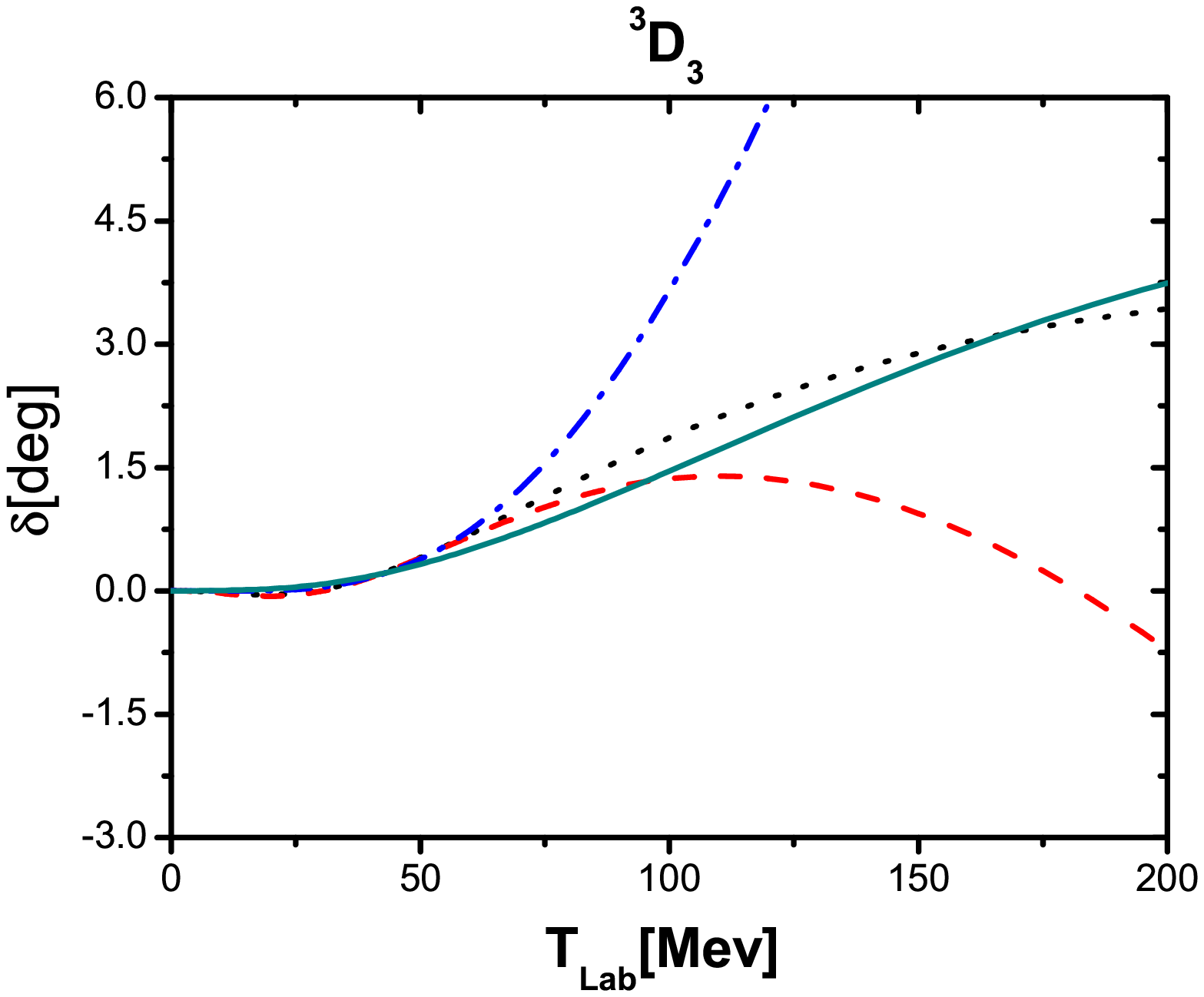}}
& \resizebox{58mm}{!}{\includegraphics{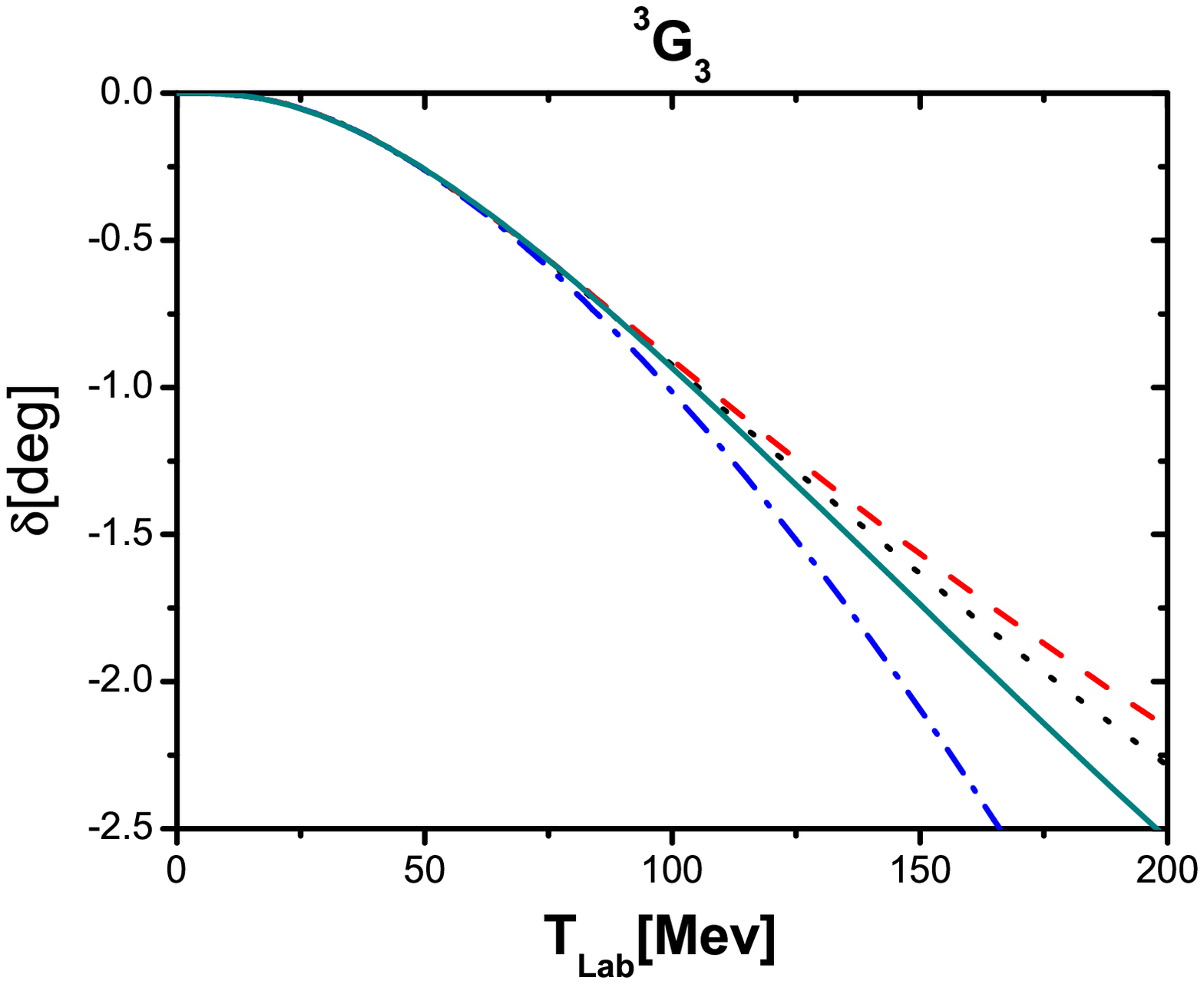}}&
     \resizebox{58mm}{!}{\includegraphics{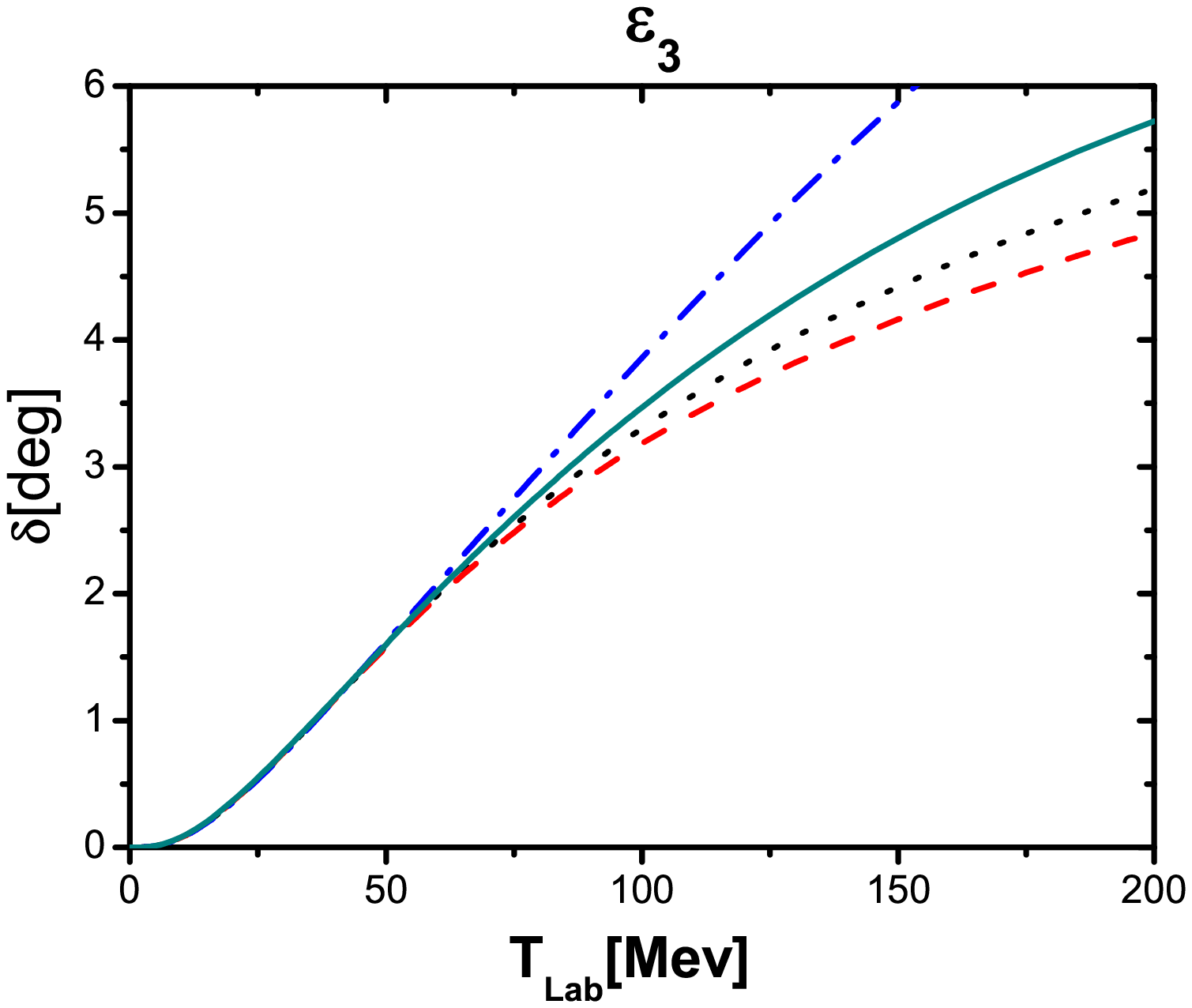}}
    \end{tabular}
   \caption{Predictions of $\delta_{^3D_3}$, $\delta_{^3G_3}$ and
   $\epsilon_3$ versus lab energy $T_{\text{Lab}}$ in MeV with the
   simplest Pad\'e, with solid line for PWA,  dotted lines for LO,
   dashed line for NLO and dot-dashed line for NNLO.}
   \label{addB}
 \end{center}
\end{figure}

\begin{figure}[h]
  \begin{center}
    \begin{tabular}{ccc}
\hspace*{-0.75cm}
     \resizebox{58mm}{!}{\includegraphics{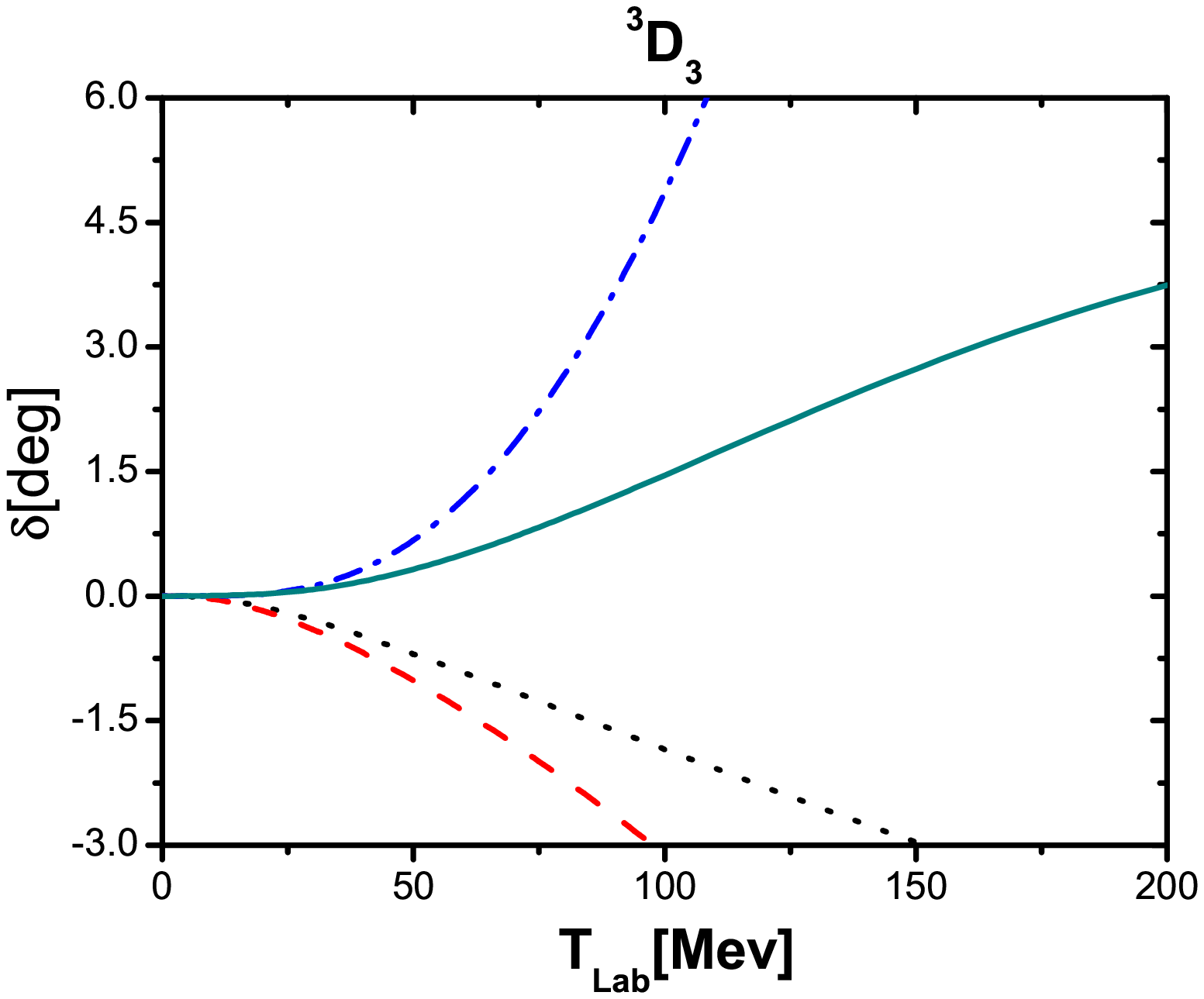}}
     &\resizebox{58mm}{!}{\includegraphics{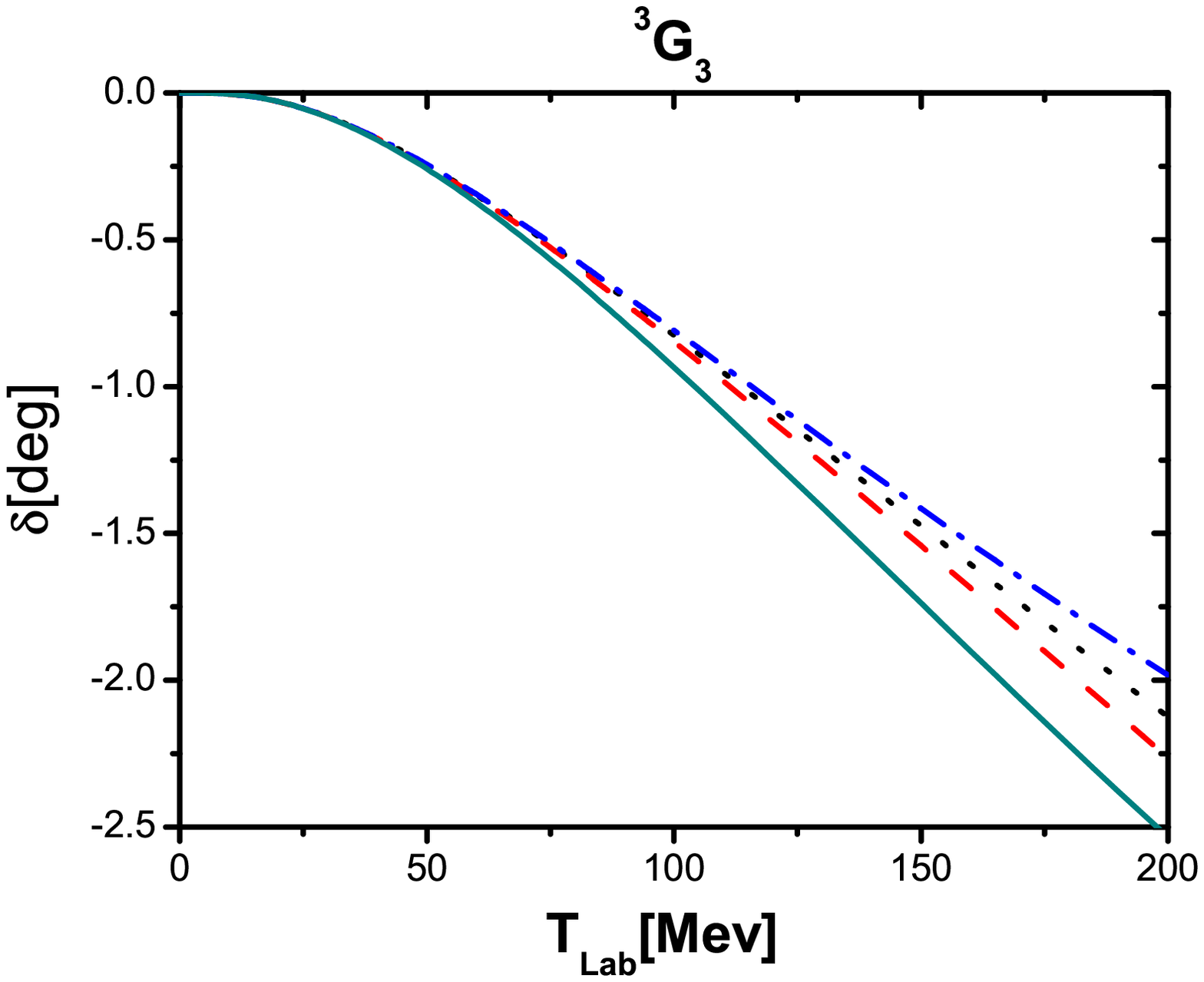}}
     &\resizebox{58mm}{!}{\includegraphics{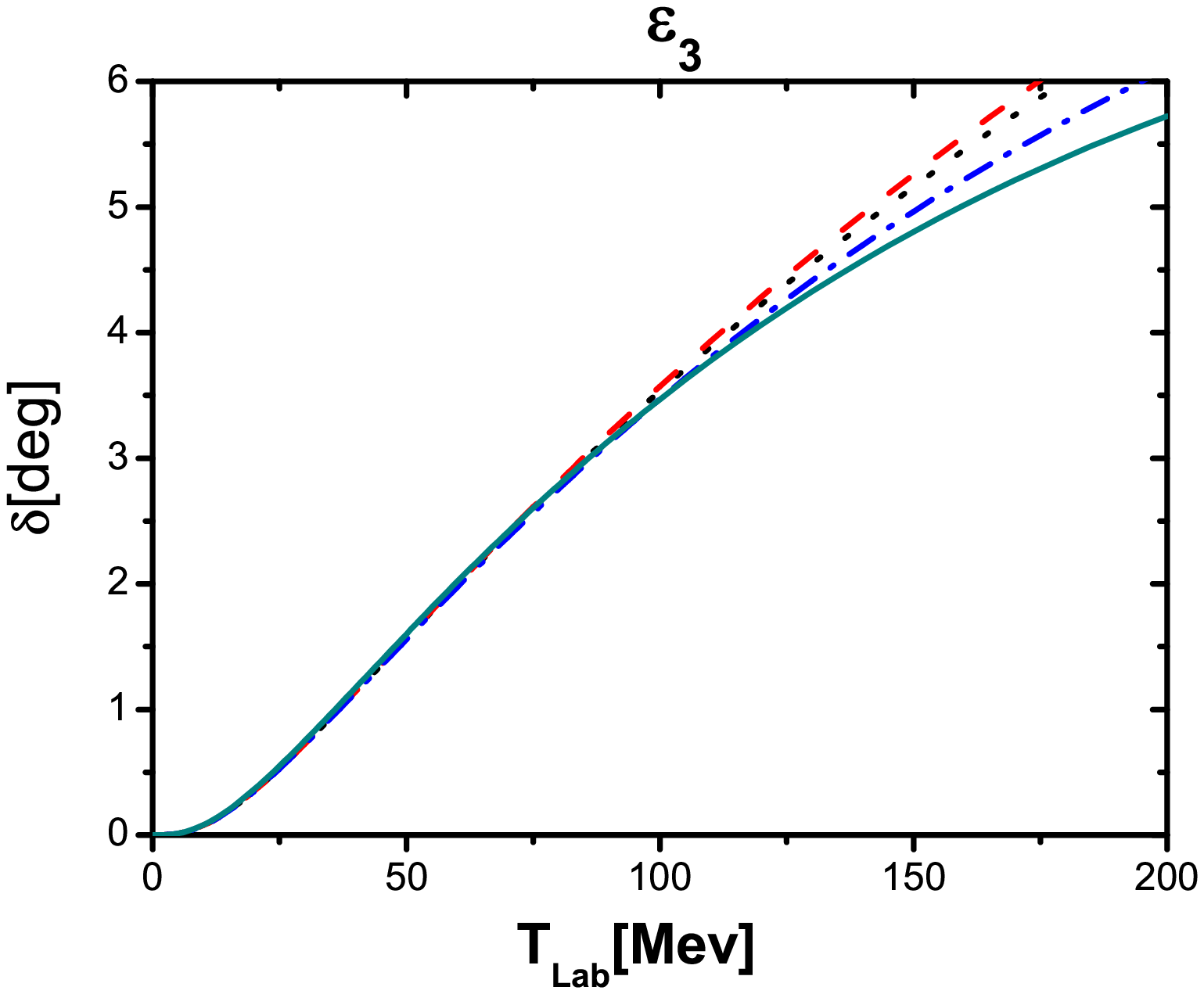}}
    \end{tabular}
   \caption{Perturbative predictions for $\delta_{^3D_3}$,
   $\delta_{^3G_3}$ and $\epsilon_3$ versus lab energy $T_{\text{Lab}}$
   in MeV, i.e., ${\mathcal{G}}$=0. Conventions are as in  Fig.\ref{addB}.}
   \label{pt}
 \end{center}
\end{figure}

The numerical results are summarized and presented in TABLE I and
Fig.\ref{addB}. The nonperturbative renormalization does
significantly improve the phase shift predictions for the $^3D_3$
channel in comparison with the inferior perturbative ones as
depicted in Fig.\ref{pt}, where the perturbative predictions even
produce wrong sign of the phase shifts. For $\delta_{^3G_3}$ we find
similar but less significant improvement for lab energies below
100MeV, as the perturbative predictions for all the three orders
already deviate from the PWA data from, say $T_{\text{Lab}}sim
75\text{MeV}$, though less significant. Such simple results already
means that only the prescription fixed through physical boundaries
could reliably describe physics. (One could try other rather
different values of $g^0_{\cdots}$ and see that the phase shifts and
mixing angle thus obtained are nonsense.) As the $p$ or energy
dependence in $\text{Re}({\mathcal{G}})$ is totally discarded here,
the predictions are doomed to fail as $E$ is higher, say,
$T_{\text{Lab}}>100\text{MeV}$, which is evident from Fig.\ref{addB}
and Fig.\ref{pt}. Also the trend that the predictions improve order
by order is not clear here. To see this trend, more sophisticated
Pad\'e approximation and hence heavier workloads are required, which
will improve the predictions in many respects. Further works along
such lines are in progress and will be reported in the near future.
Here, we are merely content with illustrating the plausibility of
Pad\'e approximant to the factor $\text{Re}({\mathcal{G}})$.
Although the Pad\'e approximant adopted here is very coarse, the
main virtues in using such relatively more analytical and
controllable approach are still quite significant from the simple
numerical analysis given in the figures and tables.

We still need to show that the Pad\'e parameters obtained via
fitting, here, $g^0_{\cdots}$, follow the rules described above in
Eq.(\ref{Pade-scale}) or (\ref{pade-PC}). To this end, we have
computed square roots of the absolute values of $g^0_{\cdots}$ and
listed them in Table I. From Table I one could see that the scale
extracted from the coarse nonperturbative approximation lies between
$10$ and $200\text{MeV}$, just in the range described by
Eq.(\ref{Pade-scale}), that is, $(10, 500)\text{MeV}$. In terms of
$\alpha$, we have $\alpha\in(0.26,0.91)\subset(0,1.0)$. In other
words, through the Pad\'e approximant of the factor ${\mathcal{G}}$,
the scales involved in the nonperturbatively renormalized
$T$-matrices do not fall outside of the EFT's scope.

At this stage, we may conclude that the Pad\'e-aided approximation
to renormalized $T$-matrices also works in coupled channels.

In summary, we extended our Pad\'e-aided analysis into coupled
channels. Primary numerical analysis showed that such treatment also
works in the coupled channels. The results also exhibit that
intensive and extensive studies are needed to further develop this
promising treatment for investigating various issues in the EFT
approach to nucleon-nucleon systems.

\newpage
\begin{table}
\caption{The simplest Pad\'e parameters for
$\text{Re}({\mathcal{G}})$ (/MeV$^2$) fitted at different chiral
orders.}
\begin{center}
    \begin{tabular}{c|c|c|c|c|c|c}\hline\hline
&$g^0_{2,2}$&$\sqrt{|g^0_{2,2}|}$ &$g^0_{2,4}$&$\sqrt{|g^0_{2,4}|}$
&$g^0_{4,4}$&$\sqrt{|g^0_{4,4}|}$
\\ \hline LO&-1300&36&4200&65&-25000&160\\
\hline NLO&-1500&39&5200&72&-33000&180\\
\hline NNLO&-770&28&210&14&5200&72\\ \hline\hline
\end{tabular}
\end{center}
    \label{table1}
\end{table}
\end{document}